\documentclass[usenatbib,letterpaper]{mn2e}
\usepackage{multirow} 
\usepackage{graphicx}
\usepackage{subfigure}
\usepackage{color}
\usepackage[authoryear]{natbib}
\usepackage{names}

\usepackage[totalwidth=480pt,totalheight=680pt,layoutvoffset=0.5cm]{geometry}

\usepackage{amsfonts}
\usepackage{amssymb}

\def\eg{{\it e.g.} }

\def\ie{{\it i.e.} }
\def\etc{{\it etc} }
\def\Herschel{{\it Herschel } }

\begin{document}
\title[]{ Planetesimal driven migration as an explanation for observations of high levels of warm, exozodiacal dust}

 \author[A. Bonsor et al.]{Amy Bonsor$^{1,2}$\thanks{Email:amy.bonsor@gmail.com}, Sean N. Raymond$^{3,4}$, Jean-Charles Augereau$^1$, Chris W. Ormel$^5$\\
$^1$UJF-Grenoble 1 / CNRS-INSU, Institut de Plan\'{e}tologie et d'Astrophysique de Grenoble (IPAG) UMR 5274, Grenoble, F-38041, France  \\     
$^2$School of Physics, H. H. Wills Physics Laboratory, University of Bristol, Tyndall Avenue, Bristol BS8 1TL, UK \\
$^3$CNRS, UMR 5804, Laboratoire d'Astrophysique de Bordeaux, 2 rue de l'Observatoire, BP 89, F-33271 Floirac Cedex, France\\
$^4$Universit\'{e} de Bordeaux, Observatoire Aquitain des Sciences de l'Univers, 2 rue de l'Observatoire, BP 89, F-33271 Floirac Cedex, France\\
$^5$Astronomy Department, University of California, Berkeley, CA 94720, USA}
                   
   \date{Received ?, 20??; accepted ?, 20??}

\maketitle

\begin{abstract}

High levels of exozodiacal dust have been observed in the inner regions of a large fraction of main sequence stars.  Given the short lifetime of the observed small dust grains, these `exozodis' are difficult to explain, especially for old ($>100$ Myr) stars. The exozodiacal dust may be observed as excess emission in the mid-infrared, or using interferometry. We hypothesise that exozodi are sustained by planetesimals scattered by planets inwards from an outer planetesimal belt, where collision timescales are long. In this work, we use N-body simulations to show that the outwards migration of a planet into a belt, driven by the scattering of planetesimals, can increase, or sustain, the rate at which planetesimals are scattered from the outer belt to the exozodi region. We hypothesise that this increase is sufficient to sustain the observed exozodi on Gyr timescales. No correlation between observations of an outer belt and an exozodi is required for this scenario to work, as the outer belt may be too faint to detect. If planetesimal driven migration does explain the observed exozodi, this work suggests that the presence of an exozodi indicates the presence of outer planets and a planetesimal belt.


\end{abstract}

\section{Introduction}
\label{sec:intro}

Emission from dusty material is commonly observed in the outer regions of planetary systems. \Herschel observations of solar-type stars find 20\% have excess emission associated with a debris disc \citep{Eiroa2013}. This presents an increase from the $\sim16\%$ of stars found by Spitzer \citep{trilling08}.  
Spitzer detected similar excess emission, from warmer dusty material, around less than 1\% of stars at 8.5-12$\mu$m \citep{Lawler2009}. Detailed modelling of many systems find that the emission is likely to originate from multiple cold and warm components \citep[e.g.][]{Hr8799su, Churcher11}. This could be compared to our Solar System, with its dusty emission from both the Kuiper and asteroid belts, as well as zodiacal dust in the terrestrial planet region.

In general cold, outer debris discs can be explained by the steady-state collisional grinding of large planetesimals  \citep{DominikDecin03, Wyatt2002, wyattreview}. However, there exists a maximum level of dust that can be produced in steady-state \citep{Wyatt07hot}. Many of the warmer dusty systems exceed this limit \citep{Wyatt07hot,Absil06, bonsor_exozodi}. The observed small dust has a short lifetime against both radiative forces and collisions. The origin of the high levels of warm dust are not fully understood. It is these systems that are the focus of this work. The term exozodi is commonly used to refer to systems with high levels of warm dust in the inner regions, referring to the similarity in location of the dust with our Solar System's zodiacal cloud.

There are two main techniques that can be used to probe high levels of warm dust in the inner regions of planetary system, with distinct abilities to detect dust with different properties. Warm emission has been detected in the mid-infrared, for example using Spitzer/IRS/Akari/WISE, from very bright, dusty systems. Near and mid-infrared interferometry, on the other hand, can probe closer to the star and detect lower levels of hotter dust. Together these observations build a picture of complex, diverse planetary systems in which dusty emission from the inner regions is common and in many cases has a complex structure, for example the dust may be split into various different belts \citep[e.g.][]{Lebreton2013, Su2013}. In this work, we focus on improving our understanding of any system in which high levels of exozodiacal dust are observed that cannot be explained by steady-state collisional evolution.

The detection statistics resulting from the mid-infrared observations are very different to those from interferometry, mainly due to the differences in sensitivities of the two techniques and the two different populations probed. Both Spitzer and WISE found that emission from dust at 8.5-12$\mu$m \citep{Lawler2009} and $12\mu$m is rare \citep{Kennedy2013}, occuring for less than 1\% of stars (Spitzer) or less than 1 in 10,000 (WISE, specifically emission from BD+20307-like systems). Interferometry has the advantage of being able to readily distinguish emission from the star and emission external to the star. A recent survey using CHARA/FLUOR found excess emission around $18^{+9}_{-5}\%$ of a sample of 28 nearby FGK stars with ages greater than 100Myr \citep{Absil2013}. Although the constraints on the position, composition and distribution of the dusty material are poor, Bayesian analysis has found in a handful of cases that the emission most likely originates from small (less than $\mu$m) grains within 1-3AU \citep{Defrere_Vega, diFolco07,Absil06}. \cite{Ertelinprep} find a similarly high occurence rate for exozodiacal dust in H-band, using PIONIER, a visitor instrument on the VLTI.

\begin{figure*}
\includegraphics[width=0.9\textwidth]{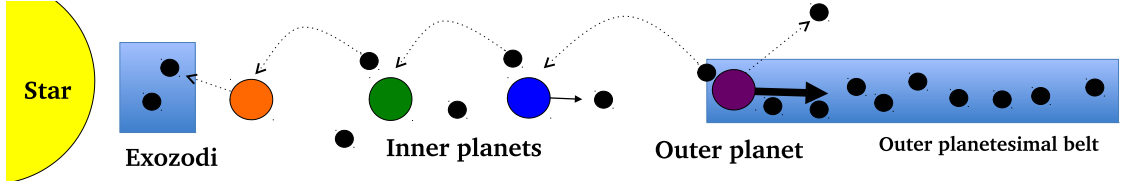}
\caption{A diagram (not to scale) to illustrate the scattering of planetesimals by an outer planet, that leads to an exchange of angular momentum and the outerward migration of that planet. Some of the scattered particles are ejected, whilst some are scattered into the inner planetary system, where they interact with the inner planets. This scattering leads to a flux of material into the exozodi region.  }
\label{fig:mig_di}
\end{figure*}

Such high levels of small grains cannot be explained in steady-state. One explanation that is regularly postulated is that we are fortunate enough to observe these systems during the aftermath of a single large collision \citep{Lisse2009,Lisse12}, maybe something like the Earth-Moon forming collision \citep{Jackson2012} or the aftermath of a dynamical instability similar to our Solar System's Late Heavy Bombardment \citep{Absil06, Lisse12,Wyatt07hot}. Not only do such collisions occur rarely at the later times ($>$100Myr) where exozodiacal dust is observed, but \cite{bonsor_instability} show that, whilst high levels of dust are produced in the inner regions of planetary systems following instabilities, this dust is short-lived, and therefore, it is highly improbable that we observe such dust in more than 0.1\% of systems. On the other hand, \citep{Kennedy2013} shows that stochastic collisions could explain the distribution of (high) excesses observed with WISE at $12\mu$m. Whilst it is not clear whether the same explanation holds for both the handful of sources with high levels of excess emission in the mid-IR, and the $\sim20\%$ of sources with K-band excess detected using interferometry \citep{Absil2013}, it is clearly easier to find an explanation for the former.

A link with outer planetary systems has been hypothesised \citep{Absil06,resolveHD69830, bonsor_exozodi, ecc_ring}. It has been shown that the emission could be reproduced in steady-state by a population of highly eccentric ($e>0.99$) planetesimals \citep{ecc_ring}, but it is not clear how such a population could form. An alternative explanation could lie in the trapping of nano grains in the magnetic field of the star \citep{Su2013, Czechowski2010}, but this still requires detailed verification. Most of the dust ($>90\%$) observed in our Solar System's zodiacal cloud is thought to originate from the disruption of Jupiter Family Comets scattered inwards by the planets from the Kuiper belt \citep{Nesvorny10}. Steady-state scattering in exo-planetary systems is not sufficiently efficient to sustain the high levels of dust observed in exozodi \citep{bonsor_exozodi}. We propose here that the efficiency at which planetesimals are scattered inwards by planets can be increased by the migration of a planet into an outer belt. We suggest that this could explain the high levels of exozodiacal dust observed in some planetary systems, in the manner illustrated by Fig.~\ref{fig:mig_di}.

In the early evolution of our Solar System, Neptune is thought to have migrated outwards, due to an exchange of angular momentum as it scattered Kuiper belt objects \citep[\eg][]{Fernandez1984,Ida2000, Levison2003,Gomes2004}. This can explain some of the observed features of the Kuiper belt, including Pluto's resonant, eccentric orbit \citep{Malhotra1993}, or the `hot classical Kuiper belt' \citep{Gomes2003}. A similar exchange of angular momentum can occur between a planet and planetesimals in an exo-planetary system \citep[\eg][]{Kirsh2009,Bromley2011, Ormel2012}. Such planetesimal driven migration can occur in either an outwards or an inwards direction, at a rate and for a timescale that depends on the properties of the disc and the planet \citep{Kirsh2009, Ormel2012}. We hypothesise that such planetesimal driven migration can sustain the scattering of planetesimals into the inner regions of a planetary system on long timescales. Such scattered planetesimals could account for the high levels of exozodiacal dust observed in some planetary systems, particularly in old ($>100$Myr) planetary systems.

In this work we use N-body simulations to consider when planetesimal driven migration occurs and its effect on the planetary system, in particular the rate at which planetesimals are scattered into the inner regions. We consider that such scattered material has the potential to resupply an exozodi on long timescales. We investigate whether the rate at which material is transported inwards is sufficient to sustain the observed exozodi. We start in \S\ref{sec:simulations} by discussing our simulations. In \S\ref{sec:migration} we discuss the migration of the outer planet in our simulations and in \S\ref{sec:scattering} the manner in which this migration increases the rate at which material is scattered into the inner regions of the planetary system. In \S\ref{sec:coll} and \S\ref{sec:observations} we discuss the collision evolution and observations of the outer disc. Our conclusions are made in \S\ref{sec:discussion}. 

\begin{table*}

\begin{tabular}{|c |c| c|}

\hline
{\bf Planet} & & \\ \hline
Outer planet semi-major axis& $a_{pl, out}$& 15AU \\
Planet semi-major axes & $a_{pl}$ & 5,6.6,8.7,11.4,15AU\\
Planet semi-major axes$^*$ & $a_{pl}$ & 5,8.7,15AU \\
Planet masses $M_{belt}(0)=10M_\oplus$ & $M_{pl}$& 2, 5, 10, 15, 20, $30M_\oplus$ \\
Planet masses $M_{belt}(0)=100M_\oplus$ & $M_{pl}$& 50, 70, 100, 130, 165, $465M_\oplus$ \\
Planet density ($M_{pl}<0.5M_\oplus$) & $\rho$& $3.9$gcm$^{-3}$\\
Planet density ($0.5<M_{pl}<10M_\oplus$) & $\rho$& 5.5gcm$^{-3}$\\
Planet density ($11<M_{pl}<49M_\oplus$)& $\rho$ &1.6gcm$^{-3}$\\
Planet density ($50<M_{pl}<500M_\oplus$) &$\rho$&1.3gcm$^{-3}$\\
Ratio between planet periods & $\frac{a_2}{a_1}$& 1.3, 1.5,1.7\\
Inner pair of planets  &\multicolumn{2}{c}{$[10M_\oplus,5M_\oplus]$, $[30M_\oplus,5M_\oplus]$, $[100M_\oplus,5M_\oplus]$, $[100M_\oplus,30M_\oplus]$, $[30M_\oplus,100M_\oplus]$}\\
& & \\
\hline
{\bf Planetesimal belt} & & \\ \hline
No. of particles & N & 2,500\\
Planetesimal mass & {\tt no mass} & 0 \\
Planetesimal mass & {\tt mass} & $M_{belt}(0)/N$\\
Total mass & $M_{belt}(0)$ & $10$, $100M_\oplus$ \\
Surface density & $\Sigma dr$ & $\propto r^{-1}dr$ \\
Semi-major axes & $a_{pp}$ & 15-30AU \\
Eccentricity & $e_{pp}$ & 0-0.01\\
Inclinations & $i_{pp}$ & $0-0.1'$\\ 
Longitude of ascending node & $\Omega_{pp}$&$0-360^\circ$\\
Longitude of pericentre & $\Lambda_{pp}$&$0-360^\circ$\\
Free anomaly &$f_{pp}$ &$0-360^\circ$\\

\hline
{\bf Other parameters} & & \\ \hline
Stellar Mass &$M_*$& $1M_\odot$ \\
Ejection radius& $r_{out}$& 10000AU \\
Inner radius &$r_{in}$ & 3AU \\
\hline

\end{tabular}
\caption{Initial conditions of our simulations}
\begin{flushleft} $^*$ For planet masses higher than $M_{pl}= 40M_\oplus$ \end{flushleft}
\label{tab:initialcond}
\end{table*}

\section{Simulations}
\label{sec:simulations}

Observations suggest that exo-planetary systems can have a wide variety of architectures. Due to the computationally intensive nature of N-body simulations that include a large number of planetesimals (each of which is assigned a mass), we must limit ourselves to an interesting range of the available parameter space. We decide for simplicity to focus on solar-like planetary systems. For want of a better example, we follow the presumed primordial belt in the Solar System \citep{Tsiganis2005, Gomes2005}, we consider a planetesimal belt that runs from 15 to 30AU and as in our Solar System, place several planets interior to the belt. It is the outer of these planets that we envisage migrating outwards through the planetesimal belt. The interior planets are necessary for planetesimals to be scattered by this outer planet inwards to the regions where an exozodi might be observed. Fig.~\ref{fig:mig_di} illustrates a planetary system of this configuration and the manner in which an exozodi may be produced. Although an exozodi may be produced by a planetary system of a different configuration, for the purposes of this study we limit ourselves to this architecture.

 A further complication arises due to the fact that the observed exozodi results from the emission due to small dust grains, yet due to the computationally intensive nature of our N-body simulations, we can only follow the evolution of `large' planetesimals. We consider the observed small dust to be a lower limit on the total mass of material that must be transported inwards in order to sustain an exozodi. In our simulations we track the mass of planetesimals that is scattered inwards. Planetesimals scattered into the inner regions of the planetary system evolve due to a variety of processes, including collisional processing, sublimation and radiative forces, for example \cite{Nesvorny10} suggest the spontaneous disruption of comets, driven by sublimating volatiles, produces the zodiacal dust in our Solar System. We envisage that these lead to the conversion of the scattered planetesimals to the observed small dust grains, with some efficiency, but exact modelling of this problem is complex and not the focus of the current work. Thus, we use the rate at which planetesimals are scattered interior to 3AU as a proxy for the maximum rate at which the observed exozodi could be replenished. This is a reasonable approximation, as long as the lifetime of the planetesimals in this region is shorter than the time bin (100Myr) that we use to count the arrival of planetesimals. We acknowledge the limitations of this technique, in particular the fact that planetesimals may be scattered onto eccentric orbits that enter and leave the region interior to 3AU without depositing significant mass.

All the initial conditions of our simulations are detailed in Table~\ref{tab:initialcond}. 
Although our simulations model the gravitational effects of the planetesimals on the planet, we neglect interactions between the planetesimals themselves. This is unlikely to be important, although it has been suggested that it could lead to some viscous spreading of the disc \citep{Ida2000, Levison2011,Raymond2012}.
In order to simplify our initial conditions we chose to place 5 equal mass planets between the pristine inner edge of the outer belt $a_{pl,out}=15$AU and $5$AU, at semi-major axes of 5, 6.6, 8.7, 11.4 and 15AU, where the ratio between adjacent pairs is constant at $\frac{a_2}{a_1}=1.3$. This purely arbitrary choice enables us to investigate the dependence on a few key parameters, for example the separation of the planets and their masses.

Our simulations were run using the hybrid integrator found in the {\it Mercury} package \citep{chambers99}.  Any planetesimals that are scattered beyond 10,000AU are considered to be ejected and we track the number of particles that are scattered within 3AU of the star. A timestep of 200 days ensures good energy conservation down to 3AU (see appendix A, \cite{Raymond2011}). 

 We run both simulations with test particles ({\tt no mass}) and simulations in which the particles are assigned a mass of $m_{part}=M_{belt}/N$, where $M_{belt}$ is the total mass of the planetesimal belt, which we initially set to either $M_{belt}(0)=10M_\oplus$ or $M_{belt}(0)=100M_\oplus$, and $N$ is the number of particles used in the simulation ($N=2,500$). The {\tt no mass} simulations provide a comparison to the {\tt mass} simulations, in which exchange of angular momentum and migration of the planets is possible. The choice of number of particles and the 1Gyr run-time for simulations were made in order to minimise the computational resources required. We note here that typical lifetimes of FGK stars can be much longer than the 1Gyr run-time of our simulations, however, given that each of the {\tt mass} simulations takes about 2 months to run for 1Gyr\footnote {On a single core on {\it fostino}, the Service Commun de Calcul Intensif de l'Observatoire de Grenoble (SCCI) supercomputer.}, we satisfy ourselves with simulating this limited portion of the system's evolution. The resolution is sufficient if the ratio of the planet mass to the particle mass is greater than 600 \citep{Kirsh2009}
, which applies to our simulations for planet masses above $2.4M_\oplus$ ($M_{belt}(0)=10M_\oplus$) or $24M_\oplus$  ($M_{belt}(0)=100M_\oplus$). This means that caution must be taken in assessing the results of the simulations with a $2M_\oplus$ planet and $M_{belt}(0)=10M_\oplus$.

\begin{figure}
\includegraphics[width=0.48\textwidth]{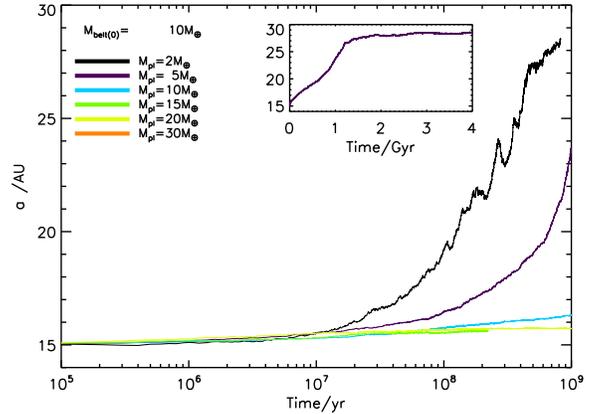}
\caption{The change in the semi-major axis of the outer planet in our simulations, for planets of various masses embedded in a disc of initial mass $10M_\oplus$. The inset shows the continued evolution of the $5M_\oplus$ planet until it reaches the outer edge of the belt. The detail in the behaviour of the $2M_\oplus$ planet is affected by the limited resolution of our simulations, particularly at late times.  Migration is stalled for planet masses greater than $10M_\oplus$.  }
\label{fig:migration_10mearth}
\end{figure}

\begin{figure*}
\includegraphics[width=0.48\textwidth]{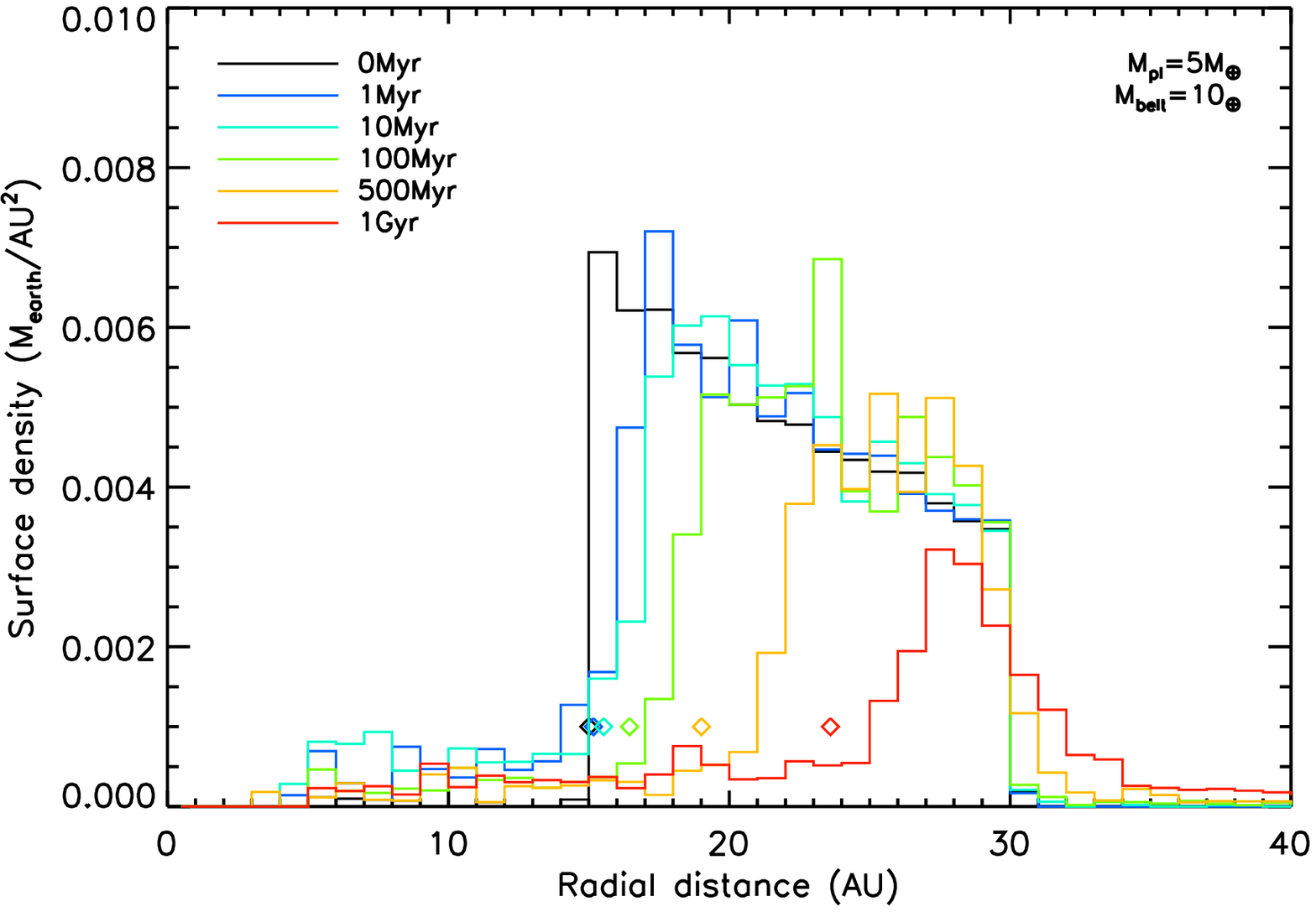}
\includegraphics[width=0.48\textwidth]{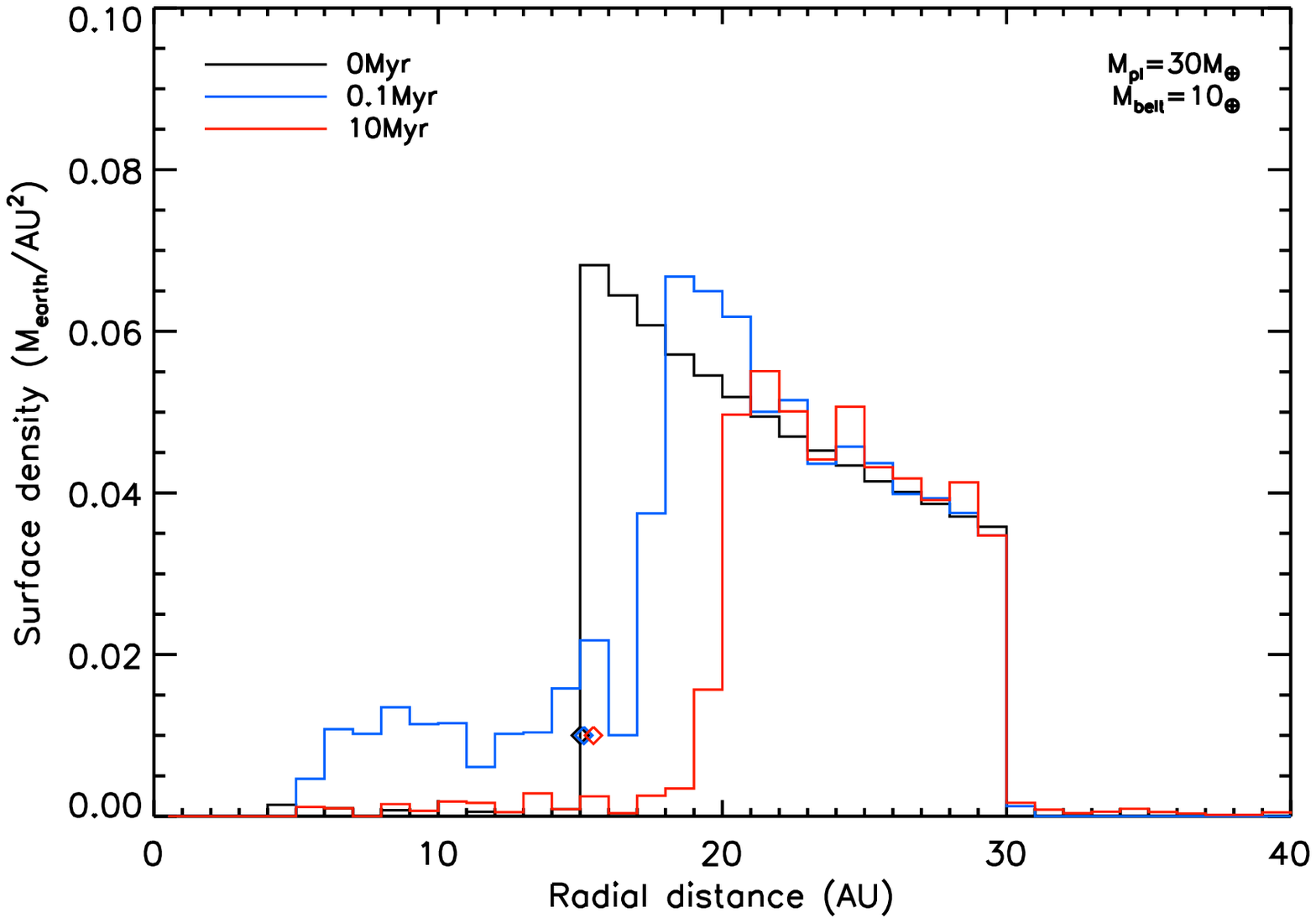}


\caption{The evolution of the surface density of planetesimals in the belt, calculated by tracking the number of planetesimals with semi-major axes in bins of width 1AU ($N_i(a)$), such that $\Sigma(a)= \Sigma_i N_i(a) m_{part}/(\pi(r_{bin,out}^2- r_{bin,in}^2))$, where $r_{bin,in}$ and $r_{bin,out}$ refer to the inner and outer edges of the bin.  The diamonds show the position of the outer planet at the different epochs. The left-hand plot shows how the $5M_\oplus$ planet clears the disc, whilst migrating through it. The right-hand plot shows how the $30M_\oplus$ planet clears the zone surrounding it of material and migrates no further. Both plots consider an initial disc mass of $10M_\oplus$. 
} 

\label{fig:surfdens}
\end{figure*}

\section{Planet migration}
\label{sec:migration}

\subsection{The conditions for migration}

 In some, but not all of our simulations, the outer planet migrates outwards. We focus initially on the simulations with an initial belt mass of $M_{belt}(0)=10M_\oplus$. Fig.~\ref{fig:migration_10mearth} shows the change in semi-major axis of the outer planet during our simulations. The migration rate depends on the relative mass of the planet to the surface density of the disc. We find three important trends in the migration rate:
\begin{itemize}
\item Migration of the outer planet is outwards
\item Migration is stalled for planets above a critical mass, $M_{stall}$
\item Lower mass planets migrate at faster rates
\end{itemize}

We are able to estimate the critical mass, $M_{stall}$ to be about $10M_\oplus$ for $M_{belt}(0)=10M_\oplus$, when the inner planets are arranged in the manner of our simulations. This mass is critically important. If migrating planets are required to produce exozodi, $M_{stall}$ tells us the maximum mass a planet can have and still migrate. The $5M_\oplus$ planet is the best example of migration that we found in our simulations. The $2M_\oplus$ planet also provides a good example of migration, but the small `blips' in its migration at later times can be accredited to the limited resolution of our simulations. 

We can readily understand the observed trends. The planet migrates due to an exchange of angular momentum as it interacts with planetesimals in a zone surrounding the planet. If these planetesimals are exterior to the planet they generally have a z-component of their specific angular momentum ($H_z=\sqrt{a(1-e^2) \cos I}$) larger than that of the planet ($H_z>H_{pl, z}$), where the subscript $_{pl}$ refers to the planet, such that interactions tend to increase the planet's angular momentum and its orbit spirals outwards \citep{Gomes2004}. On the other hand, planetesimals interior to the planet tend to have a z-component of their angular momentum lower than that of the planet ($H_z<H_{pl, z}$), such that interactions tend to decrease the planet's angular momentum and its orbit spirals inwards. At the start of our simulations, the planet has planetesimals exclusively exterior to its orbit, so in general it migrates outwards.

If we consider our best example of outwards migration, the $5M_\oplus$ planet, the left-hand panel of Fig.~\ref{fig:surfdens} illustrates how the surface density of the disc evolves as the planet migrates outwards. The planet scatters planetesimals inwards. If sufficient planetesimals were scattered inwards, the disc would be `balanced'; interactions from interior and exterior planetesimals would cancel one another. However, this scattering is not sufficient. As can be clearly seen in the figure, an imbalance in surface density remains, which `fuels' the planet's outwards migration.

Migration is not always sustained in this manner. Higher mass planets, such as the $30M_\oplus$ shown in the right-hand panel of Fig.~\ref{fig:surfdens}, scatter more planetesimals, in a potentially more `violent' manner, but tend to migrate less due to their increased inertia. This tends to `symmetrize' the disc and more importantly it carves out a deep gap in the disc around the planet, as can clearly be seen in Fig.~\ref{fig:surfdens}. This stalls the planet's migration, as the planet runs out of `fuel'.

Previous work \citep{Kirsh2009, Bromley2011, Ormel2012} has seen planet migration stall at high planet masses. \cite{Ormel2012} derive an analytic condition for a planet's migration to be stalled, by comparing the viscous stirring timescale (an approximation to the timescale to clear the planet's encounter zone of material) to the migration timescale, under the approximation of a dispersion dominated disc, such that random velocities dominate encounters, rather than Keplerian shear. This occurs for (Eq. 58 of \cite{Ormel2012} in our notation):
   \begin{equation}
M_{pl} \gtrsim M_{stall}= 23 \; \Sigma(a) \;a^2 \; e_0, 
\label{eq:mpl_stalled}
\end{equation}
where $e_0$ is the eccentricity of the planetesimals at the planet's semi-major axis, $a$ and $\Sigma(a)$ the local disc surface density. The results of our simulations confirm the existence of such a limit, although we note here that the value of $M_{stall}$ is not exact and depends on the properties of the disc, that are in turn influenced by the planet and its migration.

 Fig.~\ref{fig:mpl_sigma} illustrates diagrammatically the conditions for migration, according to Eq.~\ref{eq:mpl_stalled}, with the simplifying assumption that the disc eccentricity at the location of the planet is $e_0=0.1$, although the dashed lines indicate the variation in this line between discs with eccentricities of $e_0=0.01$ and $e_0=0.5$.

An additional regime in which very low mass planets migrate, but have little influence on the disc is also plotted on Fig.~\ref{fig:mpl_sigma}. This regime was not probed by our simulations due to our limited resolution. An analytic condition for this is derived in \cite{Ormel2012} (Eq.52 written in our notation) the minimum planet mass required to excite a noticeable eccentricity, high enough to sustain an imbalance between the disc interior and exterior to the planet:
\begin{equation} 
M_{pl} \lesssim 24 \Sigma(a) a^2 e_0^3.
\label{eq:mpl_noeffect}
\end{equation} 
Below this planet mass, we assume that the planet does not scatter sufficient planetesimals in order to produce an exozodi. Thus, whilst very low mass planets may migrate through outer discs on long timescales, they are unlikely to be the origin of exozodiacal dust.

 The initial conditions for our simulations have been added as diamonds to Fig.~\ref{fig:mpl_sigma}, which could in theory be used to predict their migration. However, we note here that the disc surface density and eccentricity will evolve as the simulation progresses, such that in general the red points move left on Fig.~\ref{fig:mpl_sigma}, whilst the upper boundary moves left and up. This means that the planet may occupy different migration regimes.  In addition, perturbations from other planets affect the migration. Thus, in practice, this plot can only be used to estimate the initial migration regime for a given planet and disc.   



\subsection{The influence of the inner planets}
\label{sec:innerpl_one}

 In order to produce an exozodi by scattering material inwards from an outer belt, several planets are required to cover the large distance between the outer belt and the exozodi. A planet that is sufficiently close to the migrating planet in order to scatter any particles that are scattered in its direction, is sufficiently close to affect the migration of the outer planet. The planet's migration is slowed by interactions with particles that it has previously scattered interior to its orbit. An inner planet can remove these particles, thus, increasing the rate, or triggering, the outer planet's migration. Inner planets can also increase the eccentricity of the planetesimals in the disc. Both of these factors can lead to substantial changes in the rate at which the outer planet migrates.

\begin{figure}
\includegraphics[width=0.48\textwidth]{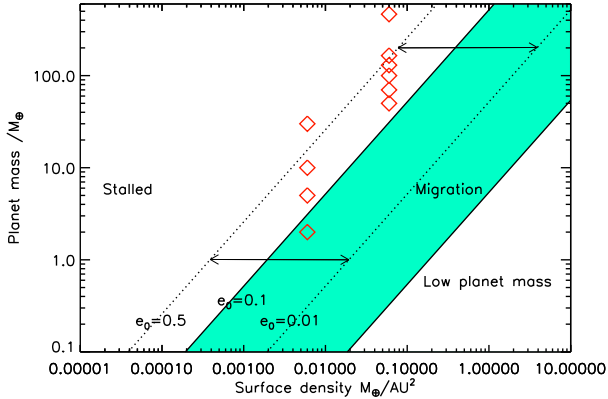}
\caption{ The range of planet masses for which the analytic criterion of \citet{Ormel2012} find that migration should occur is shown by the green shaded region (Eq.~\ref{eq:mpl_stalled} and Eq.~\ref{eq:mpl_noeffect}), assuming that the disc eccentricity is $e_0=0.1$ and $a=15$AU. The dashed lines indicates the variation in the upper boundary (Eq.~\ref{eq:mpl_stalled}) with the disc eccentricity, between $e_0=0.01$ (the start of our simulations) and $e_0=0.5$. The red diamonds indicate the initial conditions of our simulations, although it should be noted that the surface density rapidly decreases with time as the planet scatters material away. This plot should be applied with caution as the properties of the disc are liable to change due to the planet's migration and these approximations are only valid for a single planet embedded in a dispersion dominated disc.  }
\label{fig:mpl_sigma}
\end{figure}

\begin{figure}
\includegraphics[width=0.48\textwidth]{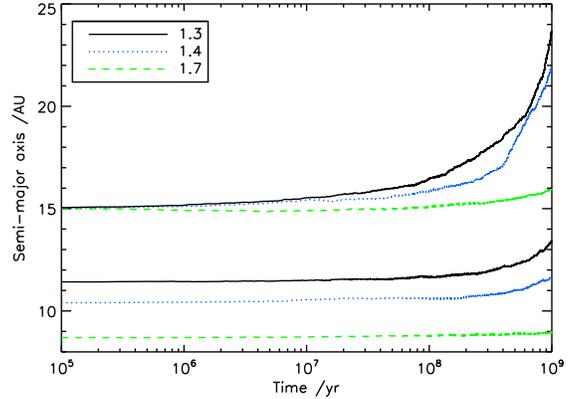}
\caption{ The change in semi-major axis of the outer two $5M_\oplus$ planets, in simulations with an initial disc mass of $10M_\oplus$, but where the initial separation of the planets is increased from $\frac{a_2}{a_1}=1.3$ (5 planets) to $\frac{a_2}{a_1}=1.4$ (4 planets) to $\frac{a_2}{a_1}=1.7$ (3 planets), as described in \S\ref{sec:simulations}. The closer the planets, the faster the migration.    }
\label{fig:a_sep}
\end{figure}

Fig.~\ref{fig:a_sep} shows the difference in the migration of the outer planet, when we changed the initial separation of the planets. This is for a fixed planet mass of $5M_\oplus$ and initial belt mass of $10M_\oplus$. This figure shows that migration occurs at higher rates if the planets are more closely packed. This is due to the increased influence of the inner planet. This figure illustrates a potentially important bias in our simulations. Our choice of planet separation moved the $5M_\oplus$ planet from a regime in which its migration was stalled (when the inner planets are far away or do not exist) to a regime in which its migration is self-sustained on Gyr timescales. This adds another difficulty in predicting exactly which planets migrate using Eq.~\ref{eq:mpl_stalled} or Fig.~\ref{fig:mpl_sigma}.


\subsection{Increasing the surface density of the disc}
\label{sec:highs}
\begin{figure}

\includegraphics[width=0.48\textwidth]{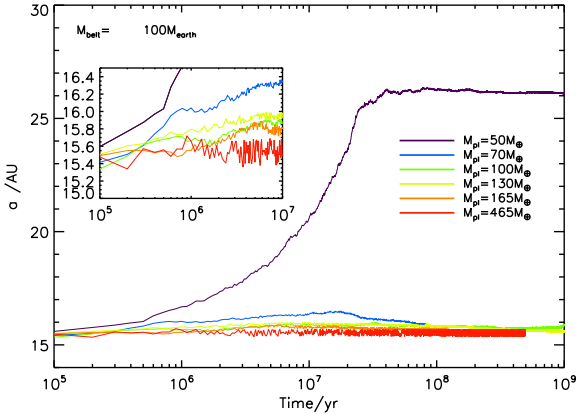}
\caption{The same as Fig.~\ref{fig:migration_10mearth}, but for an initial belt mass of $100M_\oplus$. The inset shows the first $10^7$yrs, to illustrate that most of the planets migrate, but only a small distance. Note that in order to ensure stability the inner planetary system only contained 3 planets with $\frac{a_2}{a_1}=1.7$, which has a critical influence on the migration (see discussion in \S\ref{sec:highs} for further details). Please note that if the planet was not migrating after $\sim100$Myr, due to the computationally intensive nature of the simulations, not all were evolved for 1Gyr.}
\label{fig:migration_100mearth}
\end{figure}

A similar set of simulations were run for an initial belt mass of $100M_\oplus$ (see \S\ref{sec:simulations}). According to Eq.~\ref{eq:mpl_stalled}, a higher disc surface density, means that higher mass planets will migrate. For our choice of inner planetary system with planets at 5, 6.6, 8.7, 11.4 and 15AU, with $\frac{a_2}{a_1}=1.3$, equal mass planets of greater than $40M_\oplus$ are likely to become unstable, as their separation is less than $8R_H$, where $R_H$ is the Hill's radius, given by $R_H= \frac{a_1+a_2}{2}(\frac{m_{pl,1}+m_{pl,2}}{3M_*})^{1/3} $, where $m_{pl,1}$ and $m_{pl,2}$ are the mass of the planets and $M_*$ the mass of the star. We, therefore, chose to increase the separation of the planets. This decision has a critically important effect on the migration rates, as discussed in the previous section (\S\ref{sec:innerpl_one}). This effect must be considered when analysing the migration rates of the outer planets in these simulations, shown in Fig.~\ref{fig:migration_100mearth}.

In these simulations, many of the planets initially migrate (see inset on Fig.~\ref{fig:migration_100mearth} for details, noting that the planets all started at 15AU), but their migration is commonly stalled, as the planets clear the zone surrounding them of material. Only the $50M_\oplus$ and $70M_\oplus$ planets continue to migrate on longer timescales, and then only for tens of millions of years. The migration of the $50M_\oplus$ is stalled as it reaches the outer edge of the belt. In a wider belt, this migration may well have continued on longer timescales. The migration of the $70M_\oplus$ stalls as the planet scatters particles and reduces the surface density of the disc. In fact the planet has scattered sufficient material interior to its orbit, that its migration restarts in an inwardly direction.

We note here the dependence of the migration discussed in \S\ref{sec:innerpl_one} on the architecture of the inner planetary system. Our arbitrary choice may play a critical role in our results. It may be possible to change this architecture in a manner such that migration continues on Gyr timescales even in these high surface density discs.

\subsection{The width of the belt}
In our simulations we chose to consider a belt width of 15AU. This choice is relatively arbitrary and planetesimals belts in exo-planetary systems may well be substantially wider. Migration may be sustained on longer timescales in a wider belt. Thus, the width of the belt is critically important in ascertaining whether or not migration is sustained. The $50M_\oplus$ planet in a belt of initial mass, $M_{belt}(0)=100M_\oplus$, whose migration was stalled after $\sim 30$Myr, would have continued to migrate for $\sim 100$Myr if the belt were approximately 100AU wide (and migration continued at approximately the same rate). Nonetheless, it would be difficult to sustain migration on Gyr timescales. On the other hand, the migration of the $5M_\oplus$ planet through the belt of $M_{belt}(0)=10M_\oplus$ stalls after $\sim3$Gyrs, for this width belt, somewhat less than the typical main-sequence lifetime of solar-type stars, that can be up to tens of Gyrs.

On an interesting aside, we note that for a sufficiently wide belt, the outer planet can migrate such that it becomes separated from the inner planets. In this case a stable belt may form between the two planets. The conditions for this to occur were discussed in detail in \cite{Raymond2012}.


\section{Scattering of planetesimals towards the inner planetary system }
\label{sec:scattering}
\subsection{The benefits of planet migration}
Our hypothesis is that a planet that migrates outwards continuously encounters new material in the disc that it can scatter inwards. Migrating planets, therefore, continue to scatter planetesimals on long timescales. We suppose that it is these planetesimals scattered into the inner regions of the planetary system that have the potential to replenish a dusty, exozodi. Therefore, we focus on the fate of planetesimals scattered inwards from an outer belt, for simplicity, ignoring any planetesimals that may reside initially between the planets.

We track the rate at which particles are scattered inwards, recording the first time at which a particle enters the region interior to 3AU. Scattering depends on the architecture of the inner planetary system in a complicated manner. Given that the parameter space required in order to explore this dependence fully is large, without a full investigation, our strongest conclusions result from a comparison of the scattering rates between the {\tt mass} and {\tt nomass} simulations. 
The aim of this comparison is to remove dependence on the planet masses, semi-major axes, \etc and focus solely on the difference caused by the migration of the outer planet(s). The inner planetary system does, however, have a critical effect on the migration of the outer planet (see \S\ref{sec:innerpl_one} for more details), which is hidden in this comparison.

\begin{figure}
\includegraphics[width=0.48\textwidth]{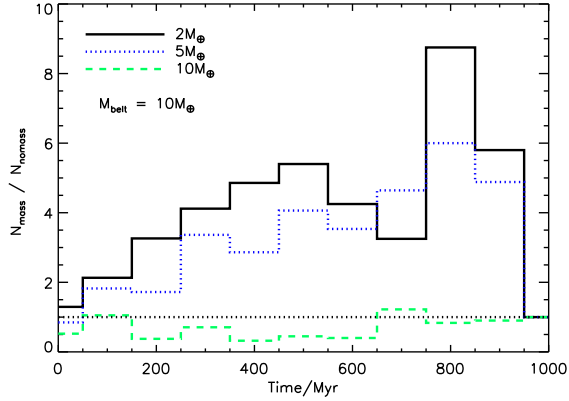}
\caption{The ratio between the number of particles scattered into the inner planetary system ($<3$AU) with ({\tt mass}) and without ({\tt nomass}) migration of the outer planet, in the simulations with an initial disc mass of $10M_\oplus$. Only those planets that migrated are shown, 2 and $5M_\oplus$, with $10M_\oplus$ for comparison. On short ($<100$Myr) timescales, the non-migrating planets scatter more material inwards, but the migrating planets continue to scatter material inwards on Gyr timescales. } 

\label{fig:ratio_10mearth}
\end{figure}

Fig.~\ref{fig:ratio_10mearth} shows the number of particles that enter the region interior to 3AU, in time bins of width 100Myr, for the {\tt mass } simulations divided by the same number for the {\tt nomass} simulations, for the low surface density outer belts ($M_{belt}(0)=10M_\oplus$). The two outer planets that migrated, with $M_{pl}=2M_\oplus$ and $5M_\oplus$, clearly scatter significantly more material inwards, than in the {\tt nomass} simulations where there was no migration. The $10M_\oplus$ whose migration was not significant, showed little difference in scattering behaviour between the {\tt mass} and {\tt nomass} simulation. This provides good evidence that migration can increase the transport of material inwards, in a manner that has the potential to produce an exozodi.

\begin{figure}

\includegraphics[width=0.48\textwidth]{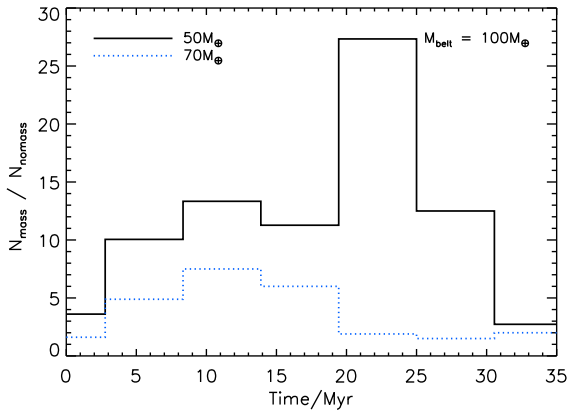}
\caption{The same as Fig.~\ref{fig:ratio_10mearth}, except for an initial disc mass of $100M_\oplus$, and a bin width of 10Myr. The $50M_\oplus$ migrated for $\sim30$Myr and the $70M_\oplus$ for $\sim 10$Myr (see Fig.~\ref{fig:migration_100mearth}), hence, the plot only shows the behaviour on these shorter timescales.  }

\label{fig:ratio_100mearth}
\end{figure}

Similar behaviour is seen for the higher surface density discs ($M_{belt}(0)=100M_\oplus$), except that here migration only continued for $<35$Myr (see Fig.~\ref{fig:ratio_100mearth}). During migration, however, the same increase in scattering rates for the {\tt mass} simulations, compared to the {\tt no mass} simulations is seen.


\begin{figure}
\includegraphics[width=0.48\textwidth]{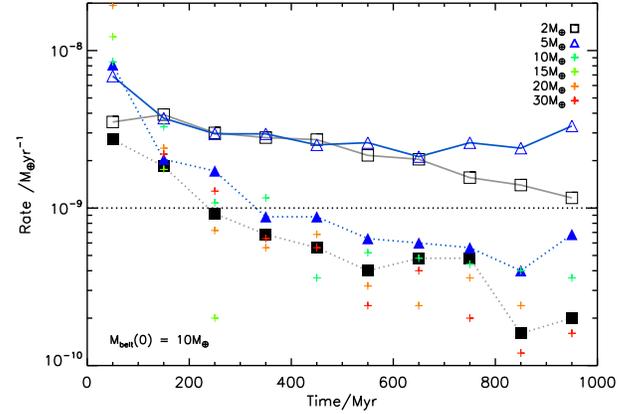}

\caption{The rate at which material is scattered interior to 3AU in the simulations with $M_{belt}(0)=10M_\oplus$, calculated by counting the number (mass) of particles that entered this region in time bins of width 100Myr. To provide an indication the $2M_\oplus$ ($5M_\oplus$) planet scattered $\sim 600$ ($\sim 800$) particles interior to 3AU, respectively. The {\tt mass} simulations where migration occurred ($M_{pl}=2,5M_\oplus$), shown by the open blue triangles and open black squares, should be compared to the simulations without migration, {\tt no mass}, shown by the filled squares and triangles. The rest of the simulations, where the outer planet's migration was insignificant, are shown by the small crosses for comparison. The horizontal line indicates a scattering rate of $10^{-9}M_\oplus yr^{-1}$, our estimate of the minimum rate required to sustain exozodi. }
\label{fig:rate}
\end{figure}


\begin{figure}
\includegraphics[width=0.48\textwidth]{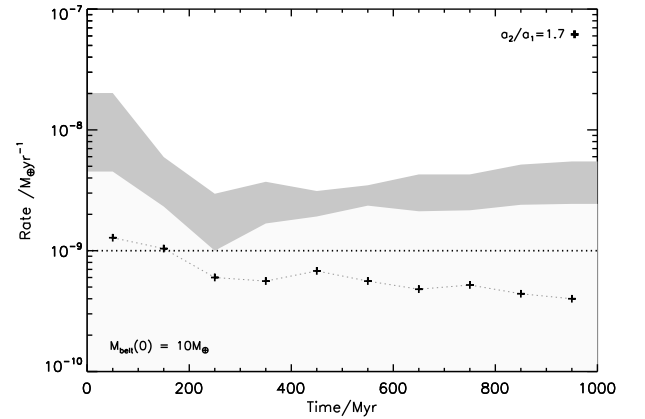}

\caption{The same as Fig.~\ref{fig:rate}, but only for an outer planet mass of $M_{pl}=5M_\oplus$, with $M_{belt}(0)=10M_\oplus$. The shaded are indicates the effect of changing the separation of the inner planets from $\frac{a_2}{a_1}= 1.3$ to $\frac{a_2}{a_1}= 1.5$,  or the masses of the inner pair of planets to $[10M_\oplus,5M_\oplus]$, $[30M_\oplus,5M_\oplus]$, $[100M_\oplus,5M_\oplus]$, $[100M_\oplus,30M_\oplus]$ and $[30M_\oplus,100M_\oplus]$. For $\frac{a_2}{a_1}=1.7$ the planet separation is so large that no planetesimals are scattered inwards. }
\label{fig:rate_changeparams}
\end{figure}


\begin{figure}
\includegraphics[width=0.48\textwidth]{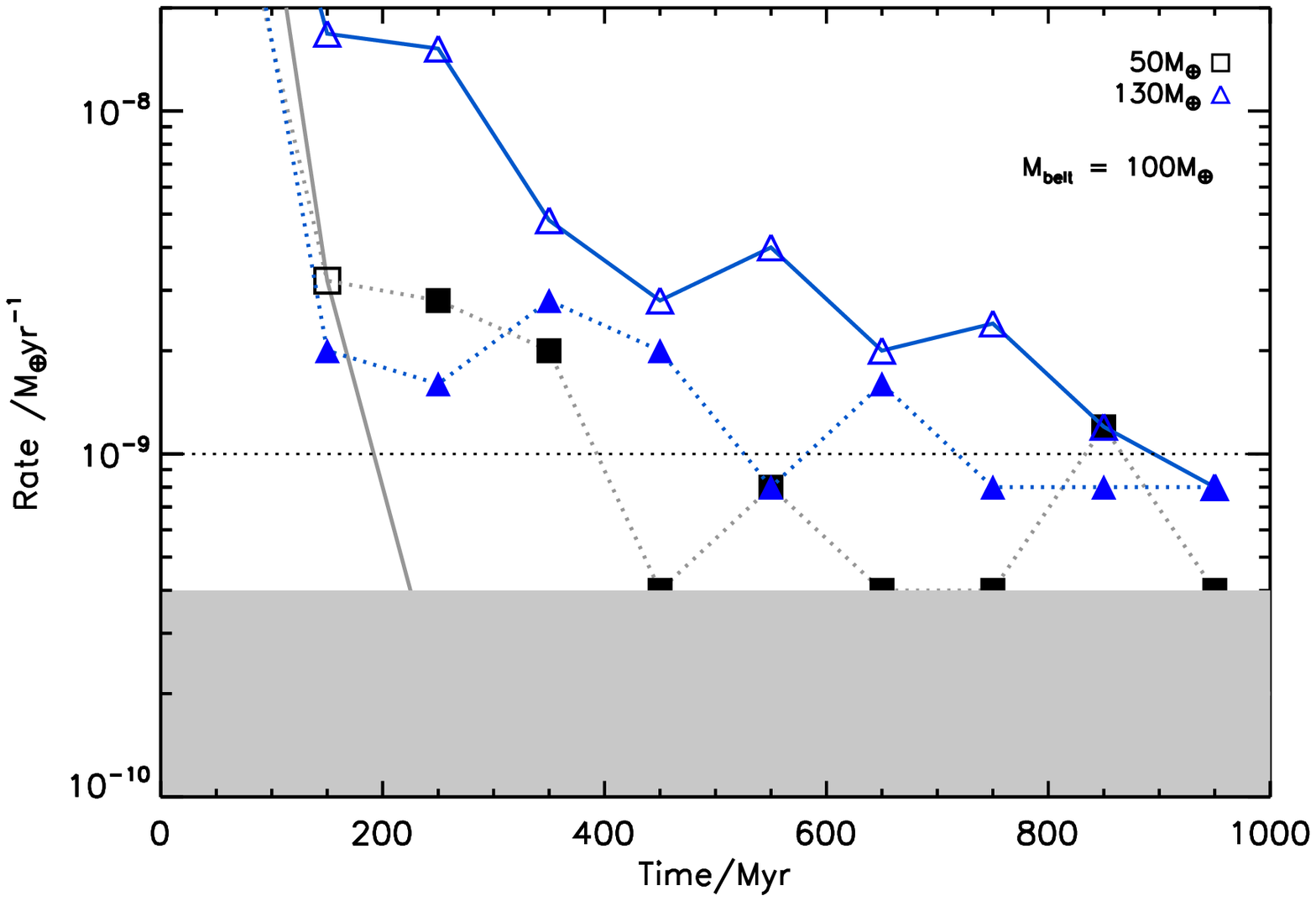}

\caption{The same as Fig.~\ref{fig:rate}, except for $M_{belt}(0)=100M_\oplus$. We show the $M_{pl}=50M_\oplus$, as example of a simulation in which the outer planet migrated and $M_{pl}=130M_\oplus$, as an example of a simulation where there was little migration of the outer planet. The solid lines and diamonds indicate the {\tt mass} simulations and the dotted lines and triangles, the {\tt nomass} simulations. The grey shaded area indicates the limit of our resolution, given by the scattering of one particle in a time bin of width 100Myr. }
\label{fig:rate_100mearth}
\end{figure}

\subsection{Is the mass flux sufficient to produce an exozodi?}
\label{sec:sufficient}
 
This is a very difficult question, given that observationally neither the mass in dust, nor its lifetime is well constrained, in addition to which, as discussed in \S\ref{sec:simulations}, the conversion of the large planetesimals scattered inwards to the small dust grains observed is complex and difficult to constrain. Firstly, considering the observations, a few examples exist where attempts have been made to estimate the mass in dust. $\tau$ Ceti is one such example, for a solar-type star. The mass of dust is estimated to be $\sim 10^{-9}M_\oplus$ in \cite{diFolco07} in grains between $0.1\mu$m and 1.5mm in size. This is estimated assuming that the dust forms a disc; the width, height, composition, size distribution and mass of which are estimated using Bayesian analysis. The collisional lifetime of this dust is estimated to be on the order of weeks to years, such that the dust must be resupplied at a rate of at least $\sim10^{-9}M_\oplus yr^{-1}$, but maybe as high as $5\times 10^{-8}M_\oplus yr^{-1}$ in order to sustain the observed levels, where the estimate should not be considered to be accurate to within more than an order of magnitude.

Secondly, we consider the exact manner in which the dust is resupplied. Collisions between larger bodies are likely to play an important role, as well as sublimation and radiative forces. The mass in planetesimals scattered inwards will be dominated by larger bodies. It may be that eventually most of this mass is converted to the observed small dust, but more likely the conversion is not 100\% efficient. In fact, it would not be unreasonable that some planetesimals scattered interior to 3AU on highly eccentric orbits, leave the inner regions of this planetary system, without depositing any mass. 

We, therefore, acknowledge that the exact mass flux required to sustain an observable exozodi is difficult to estimate and may vary from system to system. We improve our estimation of the scattering rate by using a time bin (100Myr) to count the arrival of planetesimals which is large compared to their predicted lifetime in the inner regions. We consider $\sim 10^{-9}\, M_\oplus yr^{-1}$ to be an absolute lower limit on the mass in planetesimals scattered inwards and highlight that more material may be required to sustain an observable exozodi. 

\subsubsection{Low surface density discs}
We start by plotting the rate at which planetesimals are scattered inwards in our `low density' simulations ($M_{belt}(0)=10M_\oplus$). Fig.~\ref{fig:rate} shows these rates, again calculated by counting the number of particles that enter the region interior to 3AU in time bins of width 100Myr. The two important simulations on this plot are those in which the outer planet migrated, \ie $M_{pl}=2M_\oplus$ and then $M_{pl}=5M_\oplus$. The scattering rates calculated with migration (black open squares and blue open triangles) should be compared to those in which migration did not occur (blue filled triangles and black filled squares). The critical point to take from this figure is that the migrating planets continue to scatter material inwards at rates that do not decrease significantly with time.

 As discussed above, it is difficult to estimate whether or not these scattering rates are sufficient to sustain an exozodi. The key point is that the migrating planets retain the scattering rates approximately an order of magnitude higher than without migration of the outer planet. The scattering rates for the migrating planets are above the absolute minimum mass flux required to sustain an exozodi of $10^{-9}M_\oplus yr^{-1}$. They are not, however, sufficiently higher than the absolute minimum that it is clear that these planetary systems would lead to detectable exozodi. We note here, also, that the scattering process needs to fairly efficient at transporting material inwards in order to sustain an exozodi. For 1Gyr, $1M_\oplus$ is required to sustain $10^{-9}M_\oplus yr^{-1}$, \ie 10\% of the total initial mass of this low mass belt.

\subsubsection{The effect of the inner planetary system on the scattering rates}
\label{sec:innerpl_two}

As noted in \S\ref{sec:innerpl_one}, the architecture of the inner planetary system can significantly affect the migration of the outer planet. It also has an important impact on the efficiency at which particles are scattered inwards to the exozodi region. These effects are difficult to disentangle from one another and from the results of our simulations. In \S\ref{sec:innerpl_one} we noted that the migration rates of the outer planet decrease with the separation of the innermost planets, specifically the planet directly interior to the migrating planet. This leads to a decrease in the rate at which planetesimals are scattered inwards, as shown by Fig.~\ref{fig:rate_changeparams}. In fact, with $\frac{a_2}{a_1}=1.7$, the outer planet did not migrate very far at all, and therefore, its scattering rate is similar to the {\tt no mass} simulations.

We also tested the effect of changing the mass of the innermost pair of planets (see \S\ref{sec:simulations} for details). 
This did not significantly affect the migration rates, but an increase in the masses of the innermost planets, increased the rate at which planetesimals scattered inwards reached the exozodi region ($<3$AU). Fig.~\ref{fig:rate_changeparams} shows this dependence.

\subsubsection{Higher surface density discs}

The higher surface density discs ($M_{belt}(0)=100M_\oplus$) provide an interesting comparison. Firstly, we note that a higher total disc mass, means that there is a higher mass of material that could be scattered inwards, and thus, in general, with or without migration of the outer planet, scattering rates are higher. This is because of the increased mass in the outer planet's chaotic zone. As the chaotic zone is cleared, the scattering rates fall off steeply with time. \cite{bonsor_exozodi} concluded that such steady-state scattering rates, without migration of the outer planet, are generally insufficient to sustain an exozodi. Our new results are more favourable, firstly, because the outer belt that we use is closer to the star (15AU rather 60AU) and secondly, our choice of a constant ratio of semi-major axes to separate the planets, increases the scattering rates, as resonances with the second outermost planet overlap with resonances with the outer planet in the zone just exterior to the outer planet's chaotic zone, destabilising this region.

Now, we assess whether migration of the outer planet has a significant effect on the scattering rates, as shown in Fig.~\ref{fig:rate_100mearth}. The $50M_\oplus$ planet migrated rapidly outwards, reaching the edge of the belt in $<50$Myr. The entire belt is rapidly depleted of material and the scattering rates drop off very quickly at late times.

The outwards migration of the $130M_\oplus$ in the {\tt mass} simulations, on the other hand, only lasted for a short time period and the planet moved a small distance (see inset in Fig.~\ref{fig:migration_100mearth}). This migration, however, is sufficient to increase the total reservoir of material available to the planet, as both its interior and exterior chaotic zones are now full of planetesimals. It is, therefore, able to scatter material inwards at an increased rate, on longer timescales. This explains why the scattering rates are higher for the {\tt mass} simulations than the {\tt nomass} simulations.

These results suggest that with a higher surface density disc, scattering rates might be high enough to sustain a detectable exozodi for young ($<500$Myr old) stars, but that scattering rates start to fall below the minimum around 1Gyr. This is independent of large scale migration of the outer planet, although potentially helped by initial conditions in which the outer planet's interior and exterior chaotic zones are full (as in the {\tt mass} simulation, where the planet migrated a small distance). We now note that, as discussed in the following section, collisional evolution can be important and reduce the mass available in the outer disc with time. We also question whether or not it is realistic to consider such massive discs of planetesimals, given that it is unlikely that such a massive disc did not form planets; $100M_\oplus$ of solid material, would correspond to $\sim 10,000M_\oplus \sim0.03M_\odot$ of gas, comparable to the typical disk mass inferred from sub-mm observations \citep{andrewswilliams}.


\subsubsection{Width of the belt}

The results presented here suggest that one way to maintain high scattering rates is for the outer planet to continue to migrate outwards. The outwards migration of the planet is generally stalled because the surface density of the disc reduces, either because the planet has scattered all the material away (the planets whose migration stalls initially), the disc has collisionally eroded (see \S\ref{sec:coll}) or because the planet reaches the outer edge of the belt. Self-stirring of the disc may, however, lead to the spreading of the disc over time and could in principle re-start the migration.

\begin{figure}
\includegraphics[width=0.48\textwidth]{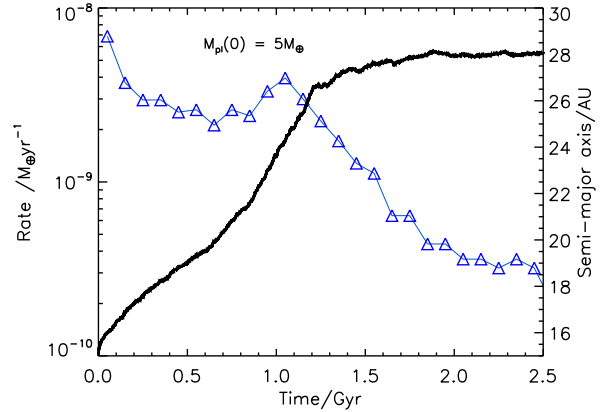}
\caption{ The scattering rates drop after the $5M_\oplus$ planet reaches the edge of the belt, with $M_{belt}(0)=10M_\oplus$. The blue triangles show the scattering rates, and the black line the outwards migration of the outer planet in this simulation.   }
\label{fig:beltwidth}
\end{figure}

Our choice of a belt width of 15AU in this work was fairly arbitrary and it is very likely that exo-planetary systems could have wider or narrower belts. Fig.~\ref{fig:beltwidth} confirms the decrease in scattering rates in one of our simulations ($M_{pl}=5M_\oplus$ and $M_{belt}(0)=10M_\oplus$), where the migration of the outer planet stopped as it reached the outer edge of the belt.

Due to the computationally intensive nature of the simulations, they were only continued for 1Gyr, yet the main-sequence lifetime of many solar-type stars can be as long as 10Gyr, and exozodi are observed around stars that are this old \citep{Absil2013}. This places another important limit regarding the ability of a planetary system to sustain an exozodi. The older the star, the wider the belt required, or the slower the migration of the outer planet, such that migration can continue on sufficiently long timescales. For our best example of the $5M_\oplus$ planet that migrated $\sim15AU$ in $\sim1.5$Gyr, assuming that migration continues at approximately the same rate, migration could continue for 5Gyrs in a 50AU wide belt. Such a wide belt, retaining the same surface density, would have a total mass of $\sim100M_\oplus$, such that only 5\% of the belt mass must be transported to the exozodi region to retain a delivery rate of $10^{-9}M_\oplus yr^{-1}$ for 5Gyrs. There is good observational evidence for the existence of such wide belts \citep[e.g.][]{bonsor_kcrb, Churcher10, muller}.

\subsubsection{Summary}

Our simulations show that sufficient mass in planetesimals can be scattered inwards from an outer planetary system, given sufficiently efficient conversion of the material scattered interior to 3AU into the observed small dust grains. In a disc with sufficiently high surface density this can occur without migration of the planets, but only for a limited time period. Fig.~\ref{fig:rate_100mearth} shows that in some cases this can be up to 500Myr-1Gyr. For a disc with a lower surface density, the outwards migration of the outer planet into an outer planetesimal belt is required. Our simulations show that this migration can be sustained for an appropriate configuration of the outer planet mass and the surface density of the disc in the region of the planet. The planet mass should not be so high that it clears the zone surrounding it of material before it has migrated across this zone, but at the same time sufficiently massive that its migration continues at a slow enough rate that it only reaches the outer edge of the belt after Gyrs. We saw this for the $5M_\oplus$ planet in our simulations with $M_{belt}(0)=10M_\oplus$. The belt must be sufficiently wide that migration is not haulted as the planet reaches its outer edge.

\section{Collisional evolution of the outer belt}
\label{sec:coll}
\subsection{The decrease in scattering rates}
Our work so far has ignored the collisional evolution of the outer belt. This is critically important. As the mass of the belt is reduced by collisions, the mass of material that can be scattered inwards decreases and it becomes harder to sustain an inwards scattering rate high enough to sustain an exozodi. In addition to this, if the disc mass decreases sufficiently, a low mass planet that was migrating through the disc, may clear its encounter zone of material faster than it migrates, such that its migration is stalled. On the other hand, collisions may themselves scatter some planetesimals into the encounter zone of the outer planets. 

A simple approximation can be used to assess the collisional evolution of the belt. We assume that it depends inversely on the collision timescale of particles in the belt, $t_c$, such that 
\begin{equation}
\frac{d M_{belt}}{dt}= -\frac{M_{belt}}{t_c}.
\label{eq:mbelt_dt}
\end{equation}
\cite{Wyatt99, wyattdent02, Wyatt07hot} derive an expression for this collisional lifetime, assuming that all bodies smaller than $D_{min}$ are removed by radiation pressure, that the mass of the disc is dominated by the largest bodies of diameter, $D_C$, and that the disc contains a Dohnanyi size distribution of bodies \citep{Dohnanyi} with a power index of $\frac{7}{2}$ \citep{Tanaka96}. It also assumes that particles can be catastrophically destroyed by particles significantly smaller than themselves, that the disc is narrow (of width $dr$), does not spread radially and that its collisional evolution started at a time $t=0$ (\ie no delayed stirring \cite[e.g.][]{alex, grantstirring}). The collisional lifetime of the disc is then given by (Eq.16 of \cite{wyattreview}) :
\begin{equation}
t_c=1.4\times 10^{-9}\;\frac{r({\scriptstyle \mathrm{AU}})^{13/3}\, (\frac{dr}{r})\, Q_D^*({\scriptstyle \mathrm{J\,kg^{-1}}})^{5/6} \, D_C({\scriptstyle \mathrm{km}) }} {M_{belt}({\scriptstyle \mathrm{M_{\oplus}}})\, M_*({\scriptstyle \mathrm{M_{\odot}}})^{4/3}\,\langle e \rangle^{5/3}} \; \; \;  {\mathrm Myr} ,
\label{eq:tc_D}
 \end{equation}
where $r$ the central radius of the belt of width $dr$, $Q_D^*$ the dispersal threshold, $D_C$ the size of the largest body in the disc, $M_*$ the stellar mass and $\langle e \rangle$ is the average eccentricity of particles in the disc, assumed to be constant. 
Values for $Q_D^*$ and $D_c$ can be determined from observations of the population of debris discs orbiting FGK stars. \cite{Kains11} find $Q_D^*=3700Jkg^{-1}$ and $D_c=450$km, with the disc eccentricity fixed at $\langle e \rangle$ fixed at $0.05$, noting the degeneracy between these parameters. Thus, the collisional lifetime of the discs considered here, with $r=22.5$AU and $dr=15$AU, is 4.2Gyr (420Myr) for belts of initial mass $10M_\oplus$ ($100M_\oplus$). In this manner the mass remaining in these discs after 1Gyr of evolution can be estimated, using $M_{belt}(t)= \frac{M_{belt}(0)}{1+ \frac{t}{t_c(0)}}$, to find $M_{belt}(1Gyr)=8M_\oplus$ ($M_{belt}(1Gyr)=30M_\oplus$).  In a similar manner after 10Gyr of evolution, $M_{belt}(10Gyr)=3M_\oplus$ ($M_{belt}(10Gyr)=4M_\oplus$).  Noting that these numbers ignore any dynamical depletion of the belt. These numbers were calculated using a very simplistic approximation, and more sophisticated models suggest that the collisional evolution may be slower than suggested here \citep[e.g.][]{lohne, Gaspar2013, Kral2013}.

As the mass in the outer belt is collisionally eroded there is less material available to be scattered inwards. Thus, it becomes more difficult to scatter sufficient material inwards to sustain an exozodi. We can consider the effect of this on our results by reducing the `mass' of each  particle, actually representative of a small sample of particles, by an appropriate factor, \ie after 1Gyr,  0.8 for the $M_{belt}=10M_\oplus$ simulations and 0.3 for the $M_{belt}=100M_\oplus$ simulations and after 10Gyr,  0.3 for the $M_{belt}=10M_\oplus$ simulations and 0.04 for the $M_{belt}=100M_\oplus$ simulations.  This is not self-consistent and does not take into account the potential reduced migration of the planet as the outer belt mass is collisionally reduced. This correction factor was not applied to Fig.~\ref{fig:rate}, but would act to decrease the scattering rates at late times. In fact, on sufficiently long timescales the disc mass will tend to the same value, regardless of the initial conditions \citep{Wyatt07hot}. This collisional evolution adds to the uncertainties in determining whether a specific scattering rate is sufficient to produce an exozodi. 

\subsection{Does collisional evolution stall the migration?}
\begin{figure}
\includegraphics[width=0.48\textwidth]{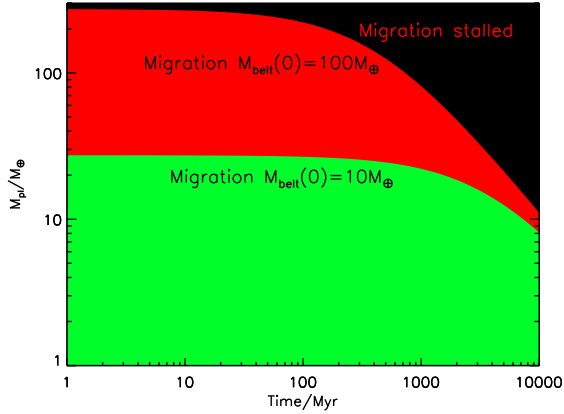}
\caption{The critical planet mass above which planetesimal driven migration stalls, calculated using Eq.~\ref{eq:mpl_stalled} and Eq.~\ref{eq:tc_D} for the evolution of the belt mass, assuming a rather excited disc with $e_0=0.5$, centred on r=22.5AU and ignoring any change in the belt width or radius.  Although the exact application of Eq.~\ref{eq:mpl_stalled} to our simulations is not clear, this plot is nonetheless indicative of the difference made by the collisional evolution of the outer belt. For the higher surface density disc, migration can be stalled relatively early due to the collisional depletion of the outer belt.  }
\label{fig:stall}
\end{figure}

In \S\ref{sec:migration} we discussed how migration stalls for planet masses above a critical mass ($M_{stall}$), that depends on the surface density (or mass) of the disc. As the disc collisionally evolves, a given planet could switch from a regime in which it migrates, to one in which its migration is stalled. We estimate where this might occur by considering the analytic formulae for $M_{stall}$, derived by \cite{Ormel2012} (Eq.~\ref{eq:mpl_stalled}), fixing the surface density of the disc at $\Sigma(a) = \frac{M_{belt}}{\pi(a_{belt,out}^2-a_{belt,in}^2)}$. Fig.~\ref{fig:stall} plots the evolution of the critical planet mass above which migration stalls, following the collisional evolution of the disc according to Eq.~\ref{eq:mbelt_dt} and Eq.~\ref{eq:tc_D}. This neglects any dynamical evolution in the disc mass, change in the disc radius or width and takes a high value for the disc eccentricity ($e_0=0.5$), in order to estimate the maximum effect of the collisional evolution. Although Eq.~\ref{eq:mpl_stalled} may not provide an exact description of the behaviour in our simulations, this plot is still indicative of the influence of collisional evolution on our simulations. Fig.~\ref{fig:stall} shows that whilst for the low surface density disc ($M_{belt}(0)=10M_\oplus$, collisional evolution is not significant and does not greatly change the migration of the outer planets, with masses less than $10M_\oplus$, that migrated in our simulations. On the other hand, for the higher surface density discs ($M_{belt}(0)=100M_\oplus$), collisional evolution can be of significant importance. For example, even in a very wide belt, the migration of a $50M_\oplus$ outer planet would be stalled after 1Gyr, merely because of the decrease in the disc surface density due to collisions, regardless of whether it reached the edge of the belt or not. A $200M_\oplus$ planet would have its migration stalled after only 100Myr.

\section{Observations of the outer belt}
\label{sec:observations}

An intriguing question, of critical importance to the interpretation of observations of exozodi, regards whether or not an outer belt must be detected in order for it to supply an exozodi. Some planetary systems have both exozodi detections and outer belts, whilst for others no outer belt is detected. \cite{Absil2013} find a tentative correlation between what they term `cold' excess and exozodi detection, or K-band excess, which they find for 4/11 FGK having previously detected outer belts, compared to 1/17 FGK stars, where there is no evidence for an outer belt. \cite{Ertelinprep}, on the other hand, find no significant correlation within their PIONIER survey, nor in a combination of the PIONIER and CHARA data. Thus, the relationship between outer belts (cold excesses) and exozodi remains unclear.

We present in this work examples of planetary systems in which the outwards migration of an outer planet, driven by the scattering of planetesimals, can sustain an exozodi. Noting that many other planetary systems may exist where this scenario could work, we focus on the results of our simulation of a $5M_\oplus$ planet migrating outwards into a belt of mass $M_{belt}(0)=10M_\oplus$. The detectability of the outer disc depends on the collisional erosion of large planetesimals that leads to the production of small dust grains. Exact predictions for this process and the detectability of a debris disc from knowledge only of the evolution of the large planetesimals, as in our case, are very difficult \citep{wyattreview, Booth09, Raymond2012}. In a similar manner it is difficult to extrapolate from the observed small dust and determine the total mass of the disc, or the size of the largest body \citep[e.g.][]{wyattreview, Greaves_tauceti, wyatt07}. Rather than making many vague approximations, we instead consider a simple approximation and track the approximate mass of material that remains in a stable disc ($e<0.02$ and $a_{pl}(1+\delta a_{chaos} )< a < 30$AU). After 1Gyr this has fallen to $1M_\oplus$, 
 not taking into account any collisional erosion of the disc.

A good example of a debris disc detected in a similar position to the belts considered in this work is $\tau$ Ceti, a nearby (3.65pc), old (10Gyr) sun-like star. The total mass of the disc (in bodies up to 10km in size) is estimated to be around $\sim1.2M_\oplus$ \citep{Greaves_tauceti}. Comparing this to our simulation, we find that our disc is still just about detectable after 1Gyr, but it is unlikely to remain detectable for long after this.  A more realistic system is unlikely to remain detectable even for this long, as collisions will erode the disc further than our naive estimate, and $\tau$ Ceti is significantly closer to the Sun that many stars with exozodi detections in \cite{Absil2013} (3.65pc, compared to up to 40pc).  We, therefore, envisage the possibility that a disc of this type continues to sustain an exozodi, without itself being hard to detect with current instruments.

Clearly these rough approximations would change depending on the exact properties of the outer belt in any given planetary system, and the exact detection properties of the instrument used to search for `cold' excess. The width of the outer belt is critically important in determining its detectability. We do not, therefore, intend to make any precise predictions regarding whether or not a detection of outer belt, makes it more likely that the exozodi is sustained by this mechanism. Instead, we merely highlight that a correlation between the presence and outer discs and exozodi is not required in order for the outer belt to sustain an exozodi.

\section{Discussion and Conclusions}
\label{sec:discussion}

High levels of dust observed in the inner regions of planetary systems, known as exozodi, are difficult to explain, particularly in old ($>100$Myr) planetary systems. In this work we investigate whether the migration of an outer planet, driven by the scattering of planetesimals, can explain these observations. We show that in principle the outwards migration of the planet maintains a reasonably high mass flux of planetesimals scattered inwards to the inner regions of the planetary systems, as long as the planet continues to migrate outwards. We hypothesise that these planetesimals could resupply the observed exozodi. An example of this scenario, applied to the specific constraints of the Vega planetary system is illustrated in \cite{RaymondVega2014}. Tight constraints on the architecture of the planetary system, however, are required in order for this scenario to work.

Firstly, we re-iterate that the constraints on how much material must be transported inwards are poor, incorporating the poor observational constraints on the mass of small dust grains, the poor theoretical constraints on the lifetime of this dust and the poorly constrained conversion of large planetesimals to small dust grains. We estimate the minimum mass flux required and show that this can be produced by planetesimals scattered inwards by planets, either if the outer planet migrates into a low surface density outer disc whilst the migration continues, or at early ($<500$Myr) times for higher surface density discs, independent of migration of the outer planet (although the efficiency of the scattering was increased in our simulations by separating the inner planets by constant ratios of their orbital periods).

We show that the migration of an outer planet can be sustained on Gyr timescales, and longer in planetary systems with wide outer belts. Whether or not a planet migrates, and how fast it migrates, depends critically on the mass of the planet and the surface density of the disc.  Migration stalls if the planet mass is sufficiently high that it clears the zone surrounding it of material before it has migrated across it. In order to sustain migration on long timescales, the best candidates are the slow migration of planets with masses just below the critical limit at which their migration is stalled, and in discs of lower surface density. The lower surface density of the disc also avoids problems with collisional depletion of the disc material stalling the migration. Wide belts are required if the migration is to continue for the full 10Gyr main-sequence lifetime of a sun-like star. This suggests that the `old' (several Gyr) planetary systems with detectable exozodis, in this scenario, would originally have had wide planetesimal belts, although these belts may by now have been cleared. It also suggests that if the frequency of planetary systems with narrow planetesimal belts is higher than that with wider planetesimal belts, there would be an increased probability of planetesimal driven migration producing exozodi around younger stars, in stark constrast to the lack of age dependence in the exozodi phenomenum found in the observations of \cite{Absil2013}. On the other hand, our estimations in \S\ref{sec:observations} suggest that the lack of a correlation between the presence of an exozodi and a detection of a cold, outer belt does not rule out the ability of planetesimal driven migration to supply the observed exozodi.

To summarise, we have presented simulations that illustrate the manner in which the outwards migration of a planet into a planetesimal belt could increase the rate at which planetesimals are scattered inwards and thus, has the potential to produce detectable exozodis, even in old ($>$Gyrs) planetary systems.

\section{Acknowledgements} 

AB acknowledges the support of the ANR-2010 BLAN-0505-01 (EXOZODI). SNR thanks the CNRS's PNP program and the ERC for their support. SNR's contribution was performed as part of the NASA Astrobiology Institute's Virtual Planetary Laboratory Lead Team, supported by the NASA under Cooperative Agreement No.  NNA13AA93A. Computations presented in this paper were performed at the Service Commun de Calcul Intensif de l'Observatoire de Grenoble (SCCI) on the super-computer funded by Agence Nationale pour la Recherche under contracts ANR-07-BLAN-0221, ANR-2010-JCJC-0504-01 and ANR-2010-JCJC-0501-01. CWO acknowledges support for this work by NASA through Hubble Fellowship grant No. HST-HF-51294.01-A awarded by the Space Telescope Science Institute, which is operated by the Association of Universities for Research in Astronomy, Inc., for NASA, under contract NAS 5-26555.

\bibliographystyle{mn}

\bibliography{ref}

\begin{thebibliography}{59}
\expandafter\ifx\csname natexlab\endcsname\relax\def\natexlab#1{#1}\fi

\bibitem[{{Absil} {et~al.}(2013){Absil}, {Defr{\`e}re}, {Coud{\'e} du Foresto},
  {Di Folco}, {M{\'e}rand}, {Augereau}, {Ertel}, {Hanot}, {Kervella},
  {Mollier}, {Scott}, {Che}, {Monnier}, {Thureau}, {Tuthill}, {ten Brummelaar},
  {McAlister}, {Sturmann}, {Sturmann}, \& {Turner}}]{Absil2013}
{Absil} O., {Defr{\`e}re} D., {Coud{\'e} du Foresto} V., {Di Folco} E.,
  {M{\'e}rand} A., {Augereau} J.-C., {Ertel} S., {Hanot} C., {Kervella} P.,
  {Mollier} B., {Scott} N., {Che} X., {Monnier} J.~D., {Thureau} N., {Tuthill}
  P.~G., {ten Brummelaar} T.~A., {McAlister} H.~A., {Sturmann} J., {Sturmann}
  L., {Turner} N., 2013, \aap, 555, A104

\bibitem[{{Absil} {et~al.}(2006){Absil}, {di Folco}, {M{\'e}rand}, {Augereau},
  {Coud{\'e} du Foresto}, {Aufdenberg}, {Kervella}, {Ridgway}, {Berger}, {ten
  Brummelaar}, {Sturmann}, {Sturmann}, {Turner}, \& {McAlister}}]{Absil06}
{Absil} O., {di Folco} E., {M{\'e}rand} A., {Augereau} J.-C., {Coud{\'e} du
  Foresto} V., {Aufdenberg} J.~P., {Kervella} P., {Ridgway} S.~T., {Berger}
  D.~H., {ten Brummelaar} T.~A., {Sturmann} J., {Sturmann} L., {Turner} N.~H.,
  {McAlister} H.~A., 2006, \aap, 452, 237

\bibitem[{{Andrews} \& {Williams}(2007)}]{andrewswilliams}
{Andrews} S.~M., {Williams} J.~P., 2007, \apj, 659, 705

\bibitem[{{Bonsor} {et~al.}(2012){Bonsor}, {Augereau}, \&
  {Th{\'e}bault}}]{bonsor_exozodi}
{Bonsor} A., {Augereau} J.-C., {Th{\'e}bault} P., 2012, \aap, 548, A104

\bibitem[{{Bonsor} {et~al.}(2013{\natexlab{a}}){Bonsor}, {Kennedy}, {Crepp},
  {Johnson}, {Wyatt}, {Sibthorpe}, \& {Su}}]{bonsor_kcrb}
{Bonsor} A., {Kennedy} G.~M., {Crepp} J.~R., {Johnson} J.~A., {Wyatt} M.~C.,
  {Sibthorpe} B., {Su} K.~Y.~L., 2013{\natexlab{a}}, \mnras, 431, 3025

\bibitem[{{Bonsor} {et~al.}(2013{\natexlab{b}}){Bonsor}, {Raymond}, \&
  {Augereau}}]{bonsor_instability}
{Bonsor} A., {Raymond} S., {Augereau} J.-C., 2013{\natexlab{b}}, \mnras,
  accepted

\bibitem[{{Booth} {et~al.}(2009){Booth}, {Wyatt}, {Morbidelli},
  {Moro-Mart{\'{\i}}n}, \& {Levison}}]{Booth09}
{Booth} M., {Wyatt} M.~C., {Morbidelli} A., {Moro-Mart{\'{\i}}n} A., {Levison}
  H.~F., 2009, \mnras, 399, 385

\bibitem[{{Bromley} \& {Kenyon}(2011)}]{Bromley2011}
{Bromley} B.~C., {Kenyon} S.~J., 2011, \apj, 735, 29

\bibitem[{{Chambers}(1999)}]{chambers99}
{Chambers} J.~E., 1999, \mnras, 304, 793

\bibitem[{{Churcher} {et~al.}(2011{\natexlab{a}}){Churcher}, {Wyatt}, \&
  {Smith}}]{Churcher10}
{Churcher} L., {Wyatt} M., {Smith} R., 2011{\natexlab{a}}, \mnras, 410, 2

\bibitem[{{Churcher} {et~al.}(2011{\natexlab{b}}){Churcher}, {Wyatt},
  {Duch{\^e}ne}, {Sibthorpe}, {Kennedy}, {Matthews}, {Kalas}, {Greaves}, {Su},
  \& {Rieke}}]{Churcher11}
{Churcher} L.~J., {Wyatt} M.~C., {Duch{\^e}ne} G., {Sibthorpe} B., {Kennedy}
  G., {Matthews} B.~C., {Kalas} P., {Greaves} J., {Su} K., {Rieke} G.,
  2011{\natexlab{b}}, \mnras, 417, 1715

\bibitem[{{Czechowski} \& {Mann}(2010)}]{Czechowski2010}
{Czechowski} A., {Mann} I., 2010, \apj, 714, 89

\bibitem[{{Defr{\`e}re} {et~al.}(2011){Defr{\`e}re}, {Absil}, {Augereau}, {di
  Folco}, {Berger}, {Coud{\'e} Du Foresto}, {Kervella}, {Le Bouquin},
  {Lebreton}, {Millan-Gabet}, {Monnier}, {Olofsson}, \& {Traub}}]{Defrere_Vega}
{Defr{\`e}re} D., {Absil} O., {Augereau} J.-C., {di Folco} E., {Berger} J.-P.,
  {Coud{\'e} Du Foresto} V., {Kervella} P., {Le Bouquin} J.-B., {Lebreton} J.,
  {Millan-Gabet} R., {Monnier} J.~D., {Olofsson} J., {Traub} W., 2011, \aap,
  534, A5

\bibitem[{{di Folco} {et~al.}(2007){di Folco}, {Absil}, {Augereau},
  {M{\'e}rand}, {Coud{\'e} du Foresto}, {Th{\'e}venin}, {Defr{\`e}re},
  {Kervella}, {ten Brummelaar}, {McAlister}, {Ridgway}, {Sturmann}, {Sturmann},
  \& {Turner}}]{diFolco07}
{di Folco} E., {Absil} O., {Augereau} J.-C., {M{\'e}rand} A., {Coud{\'e} du
  Foresto} V., {Th{\'e}venin} F., {Defr{\`e}re} D., {Kervella} P., {ten
  Brummelaar} T.~A., {McAlister} H.~A., {Ridgway} S.~T., {Sturmann} J.,
  {Sturmann} L., {Turner} N.~H., 2007, \aap, 475, 243

\bibitem[{{Dohnanyi}(1969)}]{Dohnanyi}
{Dohnanyi} J.~S., 1969, \jgr, 74, 2531

\bibitem[{{Dominik} \& {Decin}(2003)}]{DominikDecin03}
{Dominik} C., {Decin} G., 2003, \apj, 598, 626

\bibitem[{{Eiroa} {et~al.}(2013){Eiroa}, {Marshall}, {Mora}, {Montesinos},
  {Absil}, {Augereau}, {Bayo}, {Bryden}, {Danchi}, {del Burgo}, {Ertel},
  {Fridlund}, {Heras}, {Krivov}, {Launhardt}, {Liseau}, {L{\"o}hne},
  {Maldonado}, {Pilbratt}, {Roberge}, {Rodmann}, {Sanz-Forcada}, {Solano},
  {Stapelfeldt}, {Th{\'e}bault}, {Wolf}, {Ardila}, {Ar{\'e}valo}, {Beichmann},
  {Faramaz}, {Gonz{\'a}lez-Garc{\'{\i}}a}, {Guti{\'e}rrez}, {Lebreton},
  {Mart{\'{\i}}nez-Arn{\'a}iz}, {Meeus}, {Montes}, {Olofsson}, {Su}, {White},
  {Barrado}, {Fukagawa}, {Gr{\"u}n}, {Kamp}, {Lorente}, {Morbidelli},
  {M{\"u}ller}, {Mutschke}, {Nakagawa}, {Ribas}, \& {Walker}}]{Eiroa2013}
{Eiroa} C., {Marshall} J.~P., {Mora} A., {Montesinos} B., {Absil} O.,
  {Augereau} J.~C., {Bayo} A., {Bryden} G., {Danchi} W., {del Burgo} C.,
  {Ertel} S., {Fridlund} M., {Heras} A.~M., {Krivov} A.~V., {Launhardt} R.,
  {Liseau} R., {L{\"o}hne} T., {Maldonado} J., {Pilbratt} G.~L., {Roberge} A.,
  {Rodmann} J., {Sanz-Forcada} J., {Solano} E., {Stapelfeldt} K.,
  {Th{\'e}bault} P., {Wolf} S., {Ardila} D., {Ar{\'e}valo} M., {Beichmann} C.,
  {Faramaz} V., {Gonz{\'a}lez-Garc{\'{\i}}a} B.~M., {Guti{\'e}rrez} R.,
  {Lebreton} J., {Mart{\'{\i}}nez-Arn{\'a}iz} R., {Meeus} G., {Montes} D.,
  {Olofsson} G., {Su} K.~Y.~L., {White} G.~J., {Barrado} D., {Fukagawa} M.,
  {Gr{\"u}n} E., {Kamp} I., {Lorente} R., {Morbidelli} A., {M{\"u}ller} S.,
  {Mutschke} H., {Nakagawa} T., {Ribas} I., {Walker} H., 2013, \aap, 555, A11

\bibitem[{{Ertel} {et~al.}(2014){Ertel}, {Absil}, {Defr{\`e}re}, \&
  {Augereau}}]{Ertelinprep}
{Ertel} S., {Absil} O., {Defr{\`e}re} D., {Augereau} J.-C., 2014, \aap

\bibitem[{{Fernandez} \& {Ip}(1984)}]{Fernandez1984}
{Fernandez} J.~A., {Ip} W.-H., 1984, \icarus, 58, 109

\bibitem[{{G{\'a}sp{\'a}r} {et~al.}(2013){G{\'a}sp{\'a}r}, {Rieke}, \&
  {Balog}}]{Gaspar2013}
{G{\'a}sp{\'a}r} A., {Rieke} G.~H., {Balog} Z., 2013, \apj, 768, 25

\bibitem[{{Gomes} {et~al.}(2005){Gomes}, {Levison}, {Tsiganis}, \&
  {Morbidelli}}]{Gomes2005}
{Gomes} R., {Levison} H.~F., {Tsiganis} K., {Morbidelli} A., 2005, \nat, 435,
  466

\bibitem[{{Gomes}(2003)}]{Gomes2003}
{Gomes} R.~S., 2003, \icarus, 161, 404

\bibitem[{{Gomes} {et~al.}(2004){Gomes}, {Morbidelli}, \&
  {Levison}}]{Gomes2004}
{Gomes} R.~S., {Morbidelli} A., {Levison} H.~F., 2004, \icarus, 170, 492

\bibitem[{{Greaves} {et~al.}(2004){Greaves}, {Wyatt}, {Holland}, \&
  {Dent}}]{Greaves_tauceti}
{Greaves} J.~S., {Wyatt} M.~C., {Holland} W.~S., {Dent} W.~R.~F., 2004, \mnras,
  351, L54

\bibitem[{{Ida} {et~al.}(2000){Ida}, {Bryden}, {Lin}, \& {Tanaka}}]{Ida2000}
{Ida} S., {Bryden} G., {Lin} D.~N.~C., {Tanaka} H., 2000, \apj, 534, 428

\bibitem[{{Jackson} \& {Wyatt}(2012)}]{Jackson2012}
{Jackson} A.~P., {Wyatt} M.~C., 2012, \mnras, 425, 657

\bibitem[{{Kains} {et~al.}(2011){Kains}, {Wyatt}, \& {Greaves}}]{Kains11}
{Kains} N., {Wyatt} M.~C., {Greaves} J.~S., 2011, \mnras, 414, 2486

\bibitem[{{Kennedy} \& {Wyatt}(2010)}]{grantstirring}
{Kennedy} G.~M., {Wyatt} M.~C., 2010, \mnras, 405, 1253

\bibitem[{{Kennedy} \& {Wyatt}(2013)}]{Kennedy2013}
---, 2013, \mnras

\bibitem[{{Kirsh} {et~al.}(2009){Kirsh}, {Duncan}, {Brasser}, \&
  {Levison}}]{Kirsh2009}
{Kirsh} D.~R., {Duncan} M., {Brasser} R., {Levison} H.~F., 2009, \icarus, 199,
  197

\bibitem[{{Kral} {et~al.}(2013){Kral}, {Th{\'e}bault}, \& {Charnoz}}]{Kral2013}
{Kral} Q., {Th{\'e}bault} P., {Charnoz} S., 2013, \aap, 558, A121

\bibitem[{{Lawler} {et~al.}(2009){Lawler}, {Beichman}, {Bryden}, {Ciardi},
  {Tanner}, {Su}, {Stapelfeldt}, {Lisse}, \& {Harker}}]{Lawler2009}
{Lawler} S.~M., {Beichman} C.~A., {Bryden} G., {Ciardi} D.~R., {Tanner} A.~M.,
  {Su} K.~Y.~L., {Stapelfeldt} K.~R., {Lisse} C.~M., {Harker} D.~E., 2009,
  \apj, 705, 89

\bibitem[{{Lebreton} {et~al.}(2013){Lebreton}, {van Lieshout}, {Augereau},
  {Absil}, {Mennesson}, {Kama}, {Dominik}, {Bonsor}, {Vandeportal}, {Beust},
  {Defr{\`e}re}, {Ertel}, {Faramaz}, {Hinz}, {Kral}, {Lagrange}, {Liu}, \&
  {Th{\'e}bault}}]{Lebreton2013}
{Lebreton} J., {van Lieshout} R., {Augereau} J.-C., {Absil} O., {Mennesson} B.,
  {Kama} M., {Dominik} C., {Bonsor} A., {Vandeportal} J., {Beust} H.,
  {Defr{\`e}re} D., {Ertel} S., {Faramaz} V., {Hinz} P., {Kral} Q., {Lagrange}
  A.-M., {Liu} W., {Th{\'e}bault} P., 2013, \aap, 555, A146

\bibitem[{{Levison} \& {Morbidelli}(2003)}]{Levison2003}
{Levison} H.~F., {Morbidelli} A., 2003, \nat, 426, 419

\bibitem[{{Levison} {et~al.}(2011){Levison}, {Morbidelli}, {Tsiganis},
  {Nesvorn{\'y}}, \& {Gomes}}]{Levison2011}
{Levison} H.~F., {Morbidelli} A., {Tsiganis} K., {Nesvorn{\'y}} D., {Gomes} R.,
  2011, \aj, 142, 152

\bibitem[{{Lisse} {et~al.}(2009){Lisse}, {Chen}, {Wyatt}, {Morlok}, {Song},
  {Bryden}, \& {Sheehan}}]{Lisse2009}
{Lisse} C.~M., {Chen} C.~H., {Wyatt} M.~C., {Morlok} A., {Song} I., {Bryden}
  G., {Sheehan} P., 2009, \apj, 701, 2019

\bibitem[{{Lisse} {et~al.}(2012){Lisse}, {Wyatt}, {Chen}, {Morlok}, {Watson},
  {Manoj}, {Sheehan}, {Currie}, {Thebault}, \& {Sitko}}]{Lisse12}
{Lisse} C.~M., {Wyatt} M.~C., {Chen} C.~H., {Morlok} A., {Watson} D.~M.,
  {Manoj} P., {Sheehan} P., {Currie} T.~M., {Thebault} P., {Sitko} M.~L., 2012,
  \apj, 747, 93

\bibitem[{{L{\"o}hne} {et~al.}(2008){L{\"o}hne}, {Krivov}, \&
  {Rodmann}}]{lohne}
{L{\"o}hne} T., {Krivov} A.~V., {Rodmann} J., 2008, \apj, 673, 1123

\bibitem[{{Malhotra}(1993)}]{Malhotra1993}
{Malhotra} R., 1993, \nat, 365, 819

\bibitem[{{M{\"u}ller} {et~al.}(2010){M{\"u}ller}, {L{\"o}hne}, \&
  {Krivov}}]{muller}
{M{\"u}ller} S., {L{\"o}hne} T., {Krivov} A.~V., 2010, \apj, 708, 1728

\bibitem[{{Mustill} \& {Wyatt}(2009)}]{alex}
{Mustill} A.~J., {Wyatt} M.~C., 2009, \mnras, 399, 1403

\bibitem[{{Nesvorn{\'y}} {et~al.}(2010){Nesvorn{\'y}}, {Jenniskens}, {Levison},
  {Bottke}, {Vokrouhlick{\'y}}, \& {Gounelle}}]{Nesvorny10}
{Nesvorn{\'y}} D., {Jenniskens} P., {Levison} H.~F., {Bottke} W.~F.,
  {Vokrouhlick{\'y}} D., {Gounelle} M., 2010, \apj, 713, 816

\bibitem[{{Ormel} {et~al.}(2012){Ormel}, {Ida}, \& {Tanaka}}]{Ormel2012}
{Ormel} C.~W., {Ida} S., {Tanaka} H., 2012, \apj, 758, 80

\bibitem[{{Raymond} {et~al.}(2011){Raymond}, {Armitage}, {Moro-Mart{\'{\i}}n},
  {Booth}, {Wyatt}, {Armstrong}, {Mandell}, {Selsis}, \& {West}}]{Raymond2011}
{Raymond} S.~N., {Armitage} P.~J., {Moro-Mart{\'{\i}}n} A., {Booth} M., {Wyatt}
  M.~C., {Armstrong} J.~C., {Mandell} A.~M., {Selsis} F., {West} A.~A., 2011,
  \aap, 530, A62

\bibitem[{{Raymond} {et~al.}(2012){Raymond}, {Armitage}, {Moro-Mart{\'{\i}}n},
  {Booth}, {Wyatt}, {Armstrong}, {Mandell}, {Selsis}, \& {West}}]{Raymond2012}
---, 2012, \aap, 541, A11

\bibitem[{{Raymond} \& {Bonsor}(2014)}]{RaymondVega2014}
{Raymond} S.~N., {Bonsor} A., 2014, ArXiv e-prints

\bibitem[{{Smith} {et~al.}(2009){Smith}, {Wyatt}, \& {Haniff}}]{resolveHD69830}
{Smith} R., {Wyatt} M.~C., {Haniff} C.~A., 2009, \aap, 503, 265

\bibitem[{{Su} {et~al.}(2013){Su}, {Rieke}, {Malhotra}, {Stapelfeldt},
  {Hughes}, {Bonsor}, {Wilner}, {Balog}, {Watson}, {Werner}, \&
  {Misselt}}]{Su2013}
{Su} K.~Y.~L., {Rieke} G.~H., {Malhotra} R., {Stapelfeldt} K.~R., {Hughes}
  A.~M., {Bonsor} A., {Wilner} D.~J., {Balog} Z., {Watson} D.~M., {Werner}
  M.~W., {Misselt} K.~A., 2013, \apj, 763, 118

\bibitem[{{Su} {et~al.}(2009){Su}, {Rieke}, {Stapelfeldt}, {Malhotra},
  {Bryden}, {Smith}, {Misselt}, {Moro-Martin}, \& {Williams}}]{Hr8799su}
{Su} K.~Y.~L., {Rieke} G.~H., {Stapelfeldt} K.~R., {Malhotra} R., {Bryden} G.,
  {Smith} P.~S., {Misselt} K.~A., {Moro-Martin} A., {Williams} J.~P., 2009,
  \apj, 705, 314

\bibitem[{{Tanaka} {et~al.}(1996){Tanaka}, {Inaba}, \& {Nakazawa}}]{Tanaka96}
{Tanaka} H., {Inaba} S., {Nakazawa} K., 1996, \icarus, 123, 450

\bibitem[{{Trilling} {et~al.}(2008){Trilling}, {Bryden}, {Beichman}, {Rieke},
  {Su}, {Stansberry}, {Blaylock}, {Stapelfeldt}, {Beeman}, \&
  {Haller}}]{trilling08}
{Trilling} D.~E., {Bryden} G., {Beichman} C.~A., {Rieke} G.~H., {Su} K.~Y.~L.,
  {Stansberry} J.~A., {Blaylock} M., {Stapelfeldt} K.~R., {Beeman} J.~W.,
  {Haller} E.~E., 2008, \apj, 674, 1086

\bibitem[{{Tsiganis} {et~al.}(2005){Tsiganis}, {Gomes}, {Morbidelli}, \&
  {Levison}}]{Tsiganis2005}
{Tsiganis} K., {Gomes} R., {Morbidelli} A., {Levison} H.~F., 2005, \nat, 435,
  459

\bibitem[{{Wyatt}(2008)}]{wyattreview}
{Wyatt} M.~C., 2008, \araa, 46, 339

\bibitem[{{Wyatt} {et~al.}(2010){Wyatt}, {Booth}, {Payne}, \&
  {Churcher}}]{ecc_ring}
{Wyatt} M.~C., {Booth} M., {Payne} M.~J., {Churcher} L.~J., 2010, \mnras, 402,
  657

\bibitem[{{Wyatt} \& {Dent}(2002{\natexlab{a}})}]{Wyatt2002}
{Wyatt} M.~C., {Dent} W.~R.~F., 2002{\natexlab{a}}, \mnras, 334, 589

\bibitem[{{Wyatt} \& {Dent}(2002{\natexlab{b}})}]{wyattdent02}
---, 2002{\natexlab{b}}, \mnras, 334, 589

\bibitem[{{Wyatt} {et~al.}(1999){Wyatt}, {Dermott}, {Telesco}, {Fisher},
  {Grogan}, {Holmes}, \& {Pi{\~n}a}}]{Wyatt99}
{Wyatt} M.~C., {Dermott} S.~F., {Telesco} C.~M., {Fisher} R.~S., {Grogan} K.,
  {Holmes} E.~K., {Pi{\~n}a} R.~K., 1999, \apj, 527, 918

\bibitem[{{Wyatt} {et~al.}(2007{\natexlab{a}}){Wyatt}, {Smith}, {Greaves},
  {Beichman}, {Bryden}, \& {Lisse}}]{Wyatt07hot}
{Wyatt} M.~C., {Smith} R., {Greaves} J.~S., {Beichman} C.~A., {Bryden} G.,
  {Lisse} C.~M., 2007{\natexlab{a}}, \apj, 658, 569

\bibitem[{{Wyatt} {et~al.}(2007{\natexlab{b}}){Wyatt}, {Smith}, {Su}, {Rieke},
  {Greaves}, {Beichman}, \& {Bryden}}]{wyatt07}
{Wyatt} M.~C., {Smith} R., {Su} K.~Y.~L., {Rieke} G.~H., {Greaves} J.~S.,
  {Beichman} C.~A., {Bryden} G., 2007{\natexlab{b}}, \apj, 663, 365

\end{thebibliography}

\end{document}